\documentclass[useAMS,usenatbib,twocolumn]{mnras}
\usepackage{graphicx,url,amssymb,amsmath,color,units,wasysym,epsfig,epstopdf,enumerate,tabularx, subfigure,hyperref}
 
\usepackage{newtxtext,newtxmath}
\usepackage[T1]{fontenc}
\usepackage{ae,aecompl}

\usepackage[dvipsnames]{xcolor}
\usepackage[normalem]{ulem}
\usepackage[bottom]{footmisc}

\newcommand{\citeg}[1]{\citep[e.g.,][]{#1}}
\newcommand{\grb}{GRB 211211A}
\newcommand{\bnsrate}{0.22^{+8.3}_{-0.22}}
\newcommand{\bnsplusrate}{0.71^{+26.8}_{-0.70}}
\newcommand{\nsbhrate}{0.08^{+1.7}_{-0.08}}
\newcommand{\nsbhplusrate}{0.32^{+6.7}_{-0.32}}
\newcommand{\nsbhfrac}{0.09}
\newcommand{\nsbhplusfrac}{0.35}
\newcommand{\Odds}{\mathcal{O}}

\begin{document}

\title[]{Missed opportunities: \grb{} and the case for continual gravitational-wave coverage with a single observatory.}

\author[]{
Nikhil Sarin$^{1, 2}$, Paul D. Lasky$^{3,4}$, Rowina S. Nathan$^{3,4}$ \\
$^{1}$Nordita, Stockholm University and KTH Royal Institute of Technology Hannes Alfv\'ens v\"ag 12, SE-106 91 Stockholm, Sweden\\
$^2$The Oskar Klein Centre, Department of Physics, Stockholm University, AlbaNova, SE-106 91 Stockholm, Sweden\\
$^{3}$School of Physics and Astronomy, Monash University, Clayton, VIC 3800, Australia\\
$^{4}$OzGrav: The ARC Centre of Excellence for Gravitational-wave Discovery, Clayton, VIC 3800, Australia
}

\maketitle

\begin{abstract}
    Gamma-ray burst \grb{} may have been the result of a neutron star merger at $\approx350$ Mpc. However, none of the LIGO-Virgo detectors were operating at the time. We show that the gravitational-wave signal from a \grb-like binary neutron star inspiral in the next LIGO-Virgo-KAGRA observing run (O4) would be below the conventional detection threshold, however a coincident gamma-ray burst observation would provide necessary information to claim a statistically-significant multimessenger observation.  
    We calculate that with O4 sensitivity, approximately $11\%$ of gamma-ray bursts within 600 Mpc will produce a confident association between the gravitational-wave binary neutron star inspiral signature and the prompt gamma-ray signature.
    This corresponds to a coincident detection rate of $\unit[0.22^{+8.3}_{-0.22}]{yr^{-1}}$, where the uncertainties are the 90\% confidence intervals arising from uncertainties in the absolute merger rate, beaming and jet-launching fractions.
    These increase to approximately $34\%$ and $\unit[0.71^{+26.8}_{-0.70}]{yr^{-1}}$ with proposed O5 sensitivity.
    We show that the above numbers do not depend significantly on the number of gravitational-wave observatories operating with the specific sensitivity. That is, the number of confident joint gamma-ray burst and gravitational-wave detections is only marginally improved with two or three detectors operating compared to a single detector.
    It is therefore worth considering whether one detector with sufficient sensitivity (post O4) should remain in sky-watch mode at all times to elucidate the true nature of \grb-like events, a proposal we discuss in detail.
\end{abstract}

\begin{keywords}
transients: neutron star mergers -- transients: gamma-ray bursts
\end{keywords}

\section{Introduction}
\grb{} was a long-duration gamma-ray burst associated with an optical and infrared excess consistent with a kilonova~\citep{D'Ai2021, FermiGBMTeam2021}. One possible scenario is that \grb{} was the result of a binary neutron star merger, where the prolonged high-energy gamma-ray emission was due to a long-lived millisecond magnetar merger remnant or magnetic activity in an accretion disk~\citep{rastinejad22,Gompertz2022,Troja2022, Gao2022, Suvorov2022}. \grb{} observations are also consistent with the merger of a neutron star and a low-mass black hole, with the kilonova arising from both disk and dynamical ejecta~\citep{rastinejad22,Gompertz2022}. The infrared excess could have also been the thermal emission from dust and not a signature of a kilonova~\citep{Waxman2022}, in which case \grb{} is a somewhat typical long gamma-ray burst. Additional constraints from another messenger, had they existed, would have provided the necessary information to understand the physics of \grb{}.

At the distance of \grb{} ($\approx350$ Mpc), a $\unit[1.4+1.4]{M_\odot}$ binary neutron star merger would have a three-detector network optimal signal-to-noise ratio (SNR) in the LIGO-Virgo observatories~\citep{LIGO,Virgo} of 6.3, 10.3, and 17.6 with sensitivity of the third, fourth, and fifth observing runs, respectively\footnote{Note that we cannot reproduce the signal-to-noise ratios presented in \citet{rastinejad22}, despite following their outlined description.}. A neutron star-black hole merger would be louder, with signal-to-noise ratio depending on the mass of the black hole. The conventional threshold for a confident single (network) detector gravitational-wave signal is $8$ ($12$)~\citep[e.g.,][and references therein]{abbott20_LRR}. However, these thresholds are calculated assuming no electromagnetic counterpart; it is a subtle question whether signals in association of another messenger require the same conventional threshold for being called a detection. In particular, this threshold should be smaller given both the sky location and merger time of \grb{} are well known, implying the trials factors associated with gravitational-wave searches are significantly reduced. We note that some signals in the LIGO-Virgo-KAGRA gravitational-wave catalog~\citep{gwtc3} are classified as detections despite being lower than the conventional threshold for a network detection as they satisfy false-alarm-rate and astrophysical-significance estimate thresholds.

In this work, we show that a $\unit[1.4+1.4]{M_\odot}$ binary neutron star merger at $350$ Mpc with \grb-like electromagnetic observations is detectable in a LIGO-Virgo network operating at the proposed fourth observing run sensitivity. We emphasise that such a gravitational-wave signal is below the threshold for traditional gravitational-wave discovery, however the coincident electromagnetic observations provide the necessary confidence to claim a detection\footnote{We emphasise that none of our results hinge on \grb{} actually being a neutron star merger, but instead we use this event as motivation for a potential future merger.
}. An otherwise equivalent $\unit[1.4+4.5]{M_\odot}$ neutron star-black hole merger with dimensionless spin, $a = 0.85$ (sufficient to disrupt the neutron star outside the event horizon) is detectable in the network with sensitivity of the third-observing run (O3), despite only having a network optimal signal-to-noise ratio of 10.0. 

This case study provides an opportunity to analyse potential observing scenarios as the sensitivity of gravitational-wave instruments continues to improve. For example, consider the current network of two LIGO, one Virgo, and one KAGRA~\citep{KAGRA} observatory. As the sensitivity of these instruments is being upgraded, is it best practice to upgrade only three detectors at one time with one detector always online, such as the ``astro-watch'' mode of GEO600~\citep{Willke2002}? Is it optimal to have two observatories in astro-watch mode, or should we simply minimise instrument downtime by upgrading all instruments at once? While these questions also come with fiduciary and societal implications, we can somewhat answer this question from a purely scientific perspective by considering only the multi-messenger detectability of future gravitational-wave signals.

\begin{figure*}
    \centering
    \includegraphics[width=1.95\columnwidth]{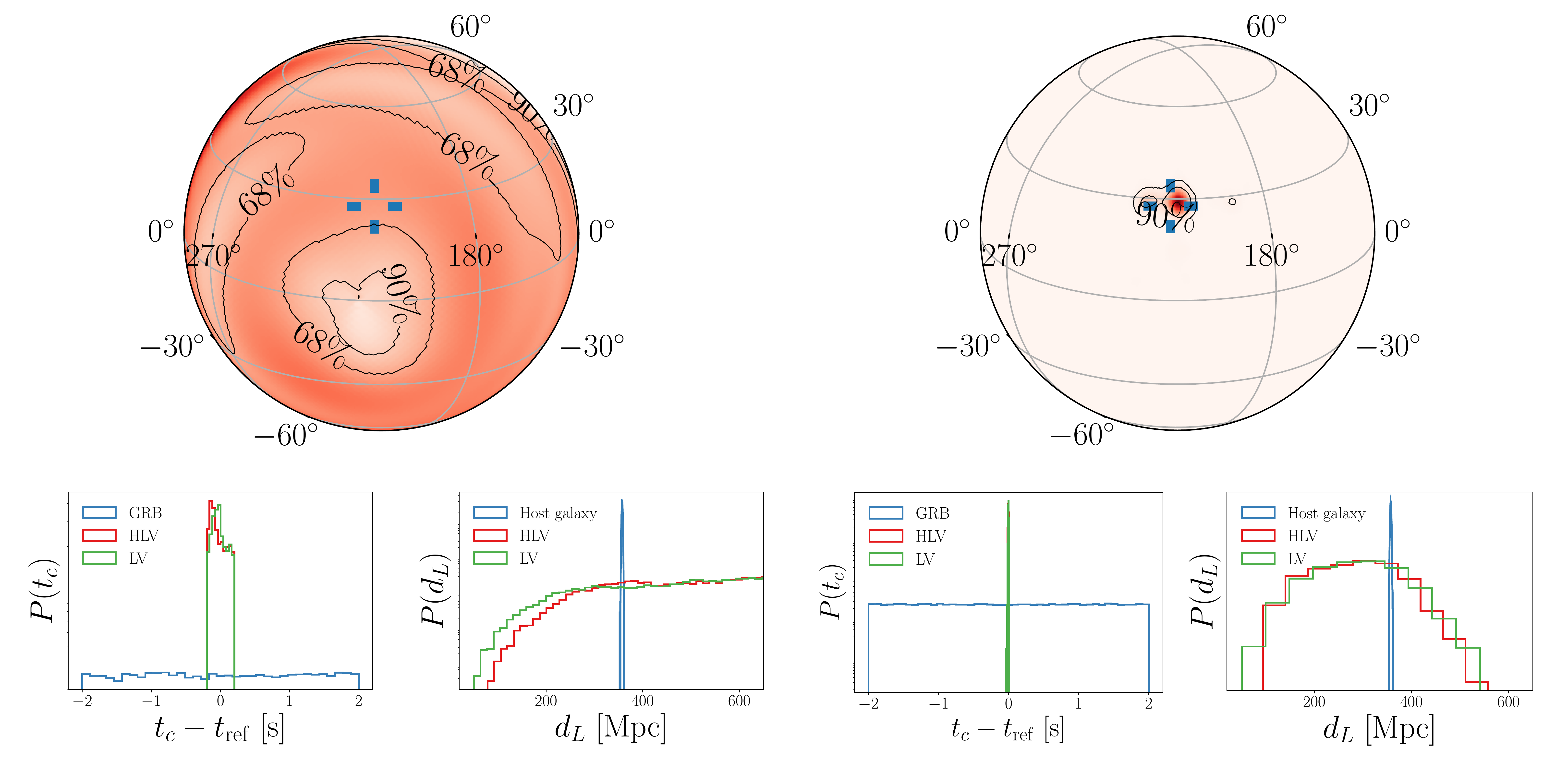}
    \caption{Posterior distributions for \grb-like injections of a binary neutron star (left) and neutron star-black hole merger into gravitational-wave networks operating at the sensitivity of the third observing run. The top panel shows the recovered posterior distributions in a three-detector network in red, and the \grb{} sky localisation from XRT on board \textit{Swift}. The one dimensional marginalised posterior distributions (bottom panels) show the coalescence time $t_c$ and luminosity distance $d_L$ assuming a three-detector network (red) and a two-detector LIGO Livingston and Virgo network (green). The blue histogram for $t_c$ assumes a four-second symmetric window around the GRB prompt emission, and for $d_L$ is derived from the host-galaxy redshift.}
    \label{fig:GRB211posteriors}
\end{figure*}

Searches for gravitational waves using only single observatories are in their infancy. Statements about detection confidence require a thorough understanding of detector noise properties because transient noise artifacts known as glitches can mimic gravitational-wave signals in single detectors~\citep[see e.g.,][]{callister17,nitz20,Cannon2021,davies22}. Detection confidence can therefore be increased by either using two or more gravitational-wave observatories searching for nearly-simultaneous and coherent signals, or by using coincident electromagnetic and/or neutrino observations to increase significance~\citep[e.g.,][]{Clark2015, nitz19,magee19,bartos19,stachie20,sarin22,Abbott2022grbb}. The latter approach increases the detection confidence of marginal gravitational-wave signals by calculating the probability that the signals originated from the same source. 

In this work, using the current best estimate for the beaming and jet-launching fractions of neutron star mergers, we predict that $\approx 11\%$ of binary neutron star mergers out to $\unit[600]{Mpc}$ will produce a joint gravitational-wave and gamma-ray burst observation with the proposed O4 sensitivity. This joint detection fraction rises to $\approx 34\%$ with the proposed O5 sensitivity. This suggests a $\unit[\bnsrate]{yr^{-1}}$ and $\unit[\bnsplusrate]{yr^{-1}}$ (90\% confidence) coincident detection rate, respectively. 
We show that the rate of joint gamma-ray burst and gravitational-wave detections is practically the same whether one, two, or three detectors remain online, providing motivation to leave a single detector online in astro-watch mode. We discuss the potential implications of such a proposal in detail. 
We note that our approach is different to other predictions of the rates of joint multi-messenger events in future gravitational-wave observing runs~\citeg{Petrov2022, Patricelli2022, Colombo2022} as we consider how confidently two independently identified events can be associated with each other, rather than a consideration of how likely an event is able to successfully detect a counterpart in follow-up to a gravitational-wave trigger. This approach relies on the conventional pipelines for serendipitous electromagnetic and gravitational-wave detection, instead of the ability to follow up a gravitational-wave sky map.


\section{\grb: The missed opportunity}
Could gravitational waves from \grb{} have been detected if the LIGO-Virgo observatories were operating at the time of the electromagnetic observations? To answer this, we inject gravitational-wave signals from a $\unit[1.4+1.4]{M_\odot}$ binary neutron star merger, and a $\unit[1.4 + 4.5]{M_\odot}$ neutron star-black hole merger, at 350 Mpc into a network of two LIGO observatories, as well as a network of two LIGO and one Virgo observatories, operating at O3 sensitivity into simulated Gaussian noise. We inject these signals at the sky location and time of \grb{}, and perform Bayesian astrophysical inference to determine how well the binary parameters, including sky localisation, coalescence time, and luminosity distance, can be determined from these mock observations.

If one observes GRB prompt emission, the system's orbital angular momentum is pointing close to the line of sight \citep[$\theta_{JN}\lesssim 10^\circ$;][]{Fong2015}. 
We perform signal injections with inclination angle $\theta_{JN}=0$; choosing a larger inclination angle consistent with electromagnetic observations will not change our results substantially. We assume zero spins for the neutron star progenitors. 
However, if the system was a neutron star-black hole binary, a $\approx 4.5\,M_\odot$ black hole would need to be spinning with dimensionless angular momentum $a\gtrsim 0.8$ to ensure tidal disruption of the neutron star~\citep{Foucart2018}, and hence the observed electromagnetic emission. We therefore assume an aligned spin for the black hole of $a=0.85$.

We perform our injections using the \texttt{Bilby} software suite~\citep{ashton19,romeroshaw20} with the \texttt{Dynesty} nested sampler~\citep{speagle20}, using standard priors on all quantities~\citep[see][]{romeroshaw20}. 
In Fig.~\ref{fig:GRB211posteriors}, we plot the recovered sky localisation posterior distributions (top rows) and $90\%$ credible interval of the coalescence time and luminosity-distance posteriors (bottom rows) for the binary neutron star merger (left) and neutron star-black hole merger (right). The sky maps show the sky-localisation posteriors for the three-detector network in red, with the blue reticle showing the sky localisation from the XRT monitor on board \textit{Swift}.
The one-dimensional marginalised coalescence time ($t_c$) and luminosity-distance ($d_L$) posteriors are shown for a three-detector network in red and a two-detector Livingston-Virgo network in green. The blue posterior distributions for the luminosity distance are from taking the host-galaxy redshift $z=0.0763\pm0.0002$~\citep{rastinejad22}, and converting to luminosity distance using Planck-18 cosmology~\citep{aghanim20}. 

The one-dimensional marginalised posterior distributions for $t_c$ and $d_L$ are largely uninformative for the binary neutron star injection into the O3 network (left-hand panels of Fig.~\ref{fig:GRB211posteriors}; note that the standard prior width for $t_c$ is \unit[0.3]{s}). On the other hand, the posteriors for the neutron star-black hole injection (right panels) are informative owing to the relatively bigger signal-to-noise ratio. For example, the posterior width on $t_c$ is approximately $\unit[0.015]{s}$.
On the other hand, the merger time as measured from the gamma/x-ray observations is less clear. For example, the gravitational-wave coalescence time for GW170817 was $\unit[1.7]{s}$ before the prompt emission~\citep{abbott17_GW170817_mma}. \grb{} had x-ray precursor emission approximately one second before the prompt~\citep{xiao22}. Taking an agnostic view towards the origin of this pre-cursor emission, we estimate the posterior for the time of coalescence (i.e., merger time) from the electromagnetic observations as being uniform over a four-second range centred around the time of onset of the prompt emission. These estimated posterior distributions are shown in blue in Fig.~\ref{fig:GRB211posteriors}. 

Determining the detectability of the gravitational-wave signal given the electromagnetic observation requires calculating the probability that the electromagnetic and gravitational-wave signals came from the same source. For this we employ the bAyesian Coincident Detection Criterion~\citep[AC/DC;][]{ashton18}.
We calculate overlap integrals of the posterior distributions using each of the separate data sets (i.e., gravitational-wave, prompt, and host galaxy identification; for details of the calculation, see~\citet{ashton18} and \citet{sarin22}). 
We then calculate the odds $\Odds_{C/R}$, where $C$ refers to the hypothesis that the independent gravitational-wave and electromagnetic data come from a common origin, and $R$ refers to the hypothesis that they are random (i.e., not associated with one another). 

The odds is the multiplication of the prior odds $\pi_{C/R}$ with each common parameter's overlap integral~\citep[see][for details]{ashton18,sarin22}. By definition, if signals come from a common event, parameters such as the masses and spins of the binary components must be the same. Unfortunately, models of short gamma-ray bursts and kilonovae that connect to the intrinsic binary parameters are not robust. Instead, we only assume the merger time, luminosity distance, and sky location are common across the gravitational-wave and electromagnetic signal. The increase in odds obtained by including any other parameters is proportional to how informed the posteriors for that parameter are with electromagnetic and gravitational-wave data relative to their prior. 

Estimating the prior odds requires understanding both the rate of gravitational-wave triggers and the rate of gamma-ray triggers~\citep[for details, see][]{ashton18}
\begin{equation}\label{eq:priorodds}
\pi_{C/R} = \frac{R_{\rm{GW,EM}}}{R_{\rm GW}R_{\rm EM} T}.
\end{equation}
Here $R_{\rm{GW,EM}}$ is the rate of astrophysical gravitational-wave and electromagnetic signals, i.e., the rate at which neutron star binaries that can produce an electromagnetic counterpart occur, $R_{\rm{GW}}$ and $R_{\rm EM}$ are the rate of gravitational-wave and electromagnetic triggers respectively, and $T$ is the co-observing time. We take $T=\unit[4]{s}$ but emphasise that this factor cancels out when multiplied by the overlap integral of the time of coalescence~\citep[see][for details]{ashton18,sarin22}.

As a rule of thumb, the false-alarm rate (i.e., the rate of triggers) in a single gravitational-wave detector goes down two orders of magnitude for every one unit of signal-to-noise ratio $\rho$; i.e., 
\begin{align}
    {\rm FAR}\propto 10^{-\xi \rho},\label{eq:FAR}
\end{align} 
with $\xi\approx{2}$~\citep{kelley13, davies22}. We use the above relation, calibrating it to the rate of single detector searches for gravitational waves from binary black holes~\citep{callister17}\footnote{See their Figure 1}. We note that this is likely an overestimate for the rate of gravitational-wave triggers, as both neutron star black hole and binary neutron star mergers are longer duration signals, and therefore less likely to be mistaken for instrumental glitches. Furthermore, having two or more detectors also allows for coherence tests making the above choice a conservative estimate for the prior odds in that scenario. 

In principle, prompt gamma-ray observations also have a similar relationship to that described above between the rate of triggers and the signal-to-noise ratio. However, as most of the gamma-ray bursts observed in the visible volume of gravitational-wave detectors will be electromagnetically bright, we can continue to use the estimate of $\unit[2]{day^{-1}}$~\citep{vonKienlin2020}. We note that \grb{} was also observed by \textit{Swift} and identification of the afterglow, kilonova and host-galaxy would increase the significance of the trigger. However, \textit{Fermi} triggers more often, implying that our prior odds calculation is conservative.

Given we know the signal-to-noise ratio of our injected signals, we can calculate the likelihood that a trigger of that signal-to-noise ratio or greater is a false alarm in our detector. We then use this number, combined with $R_{\rm EM}$ and $R_{\rm{GW,EM}}$ in Eq.~\ref{eq:priorodds} to determine the prior odds. 
To estimate $R_{\rm{GW,EM}}$, we use absolute binary neutron star and neutron star-black hole merger rates of $\unit[10-1700]{Gpc^3\,yr^{-1}}$ and $\unit[8-140]{Gpc^{-3}\,yr^{-1}}$ respectively~\citep[90\% credible intervals][]{abbott22_gwtc3pop}\footnote{We use the rate constraints marginalised over the different mass models from~\citet{abbott22_gwtc3pop}, with our median value being the median value from the binned Gaussian method in that work.}, and beaming and jet-launching fractions from~\citet{sarin22_linking}.
For example, for the binary neutron star merger injection into a three-detector network at O3 sensitivity, the network SNR is $7.2$, corresponding to a prior odds $0.01-26$ (90\% credible interval). That is, a priori it is between $0.01-26$ times more likely that a coincident gravitational-wave trigger of SNR=$7.2$ with a gamma-ray burst trigger is of astrophysical origin and from the same source. 

We set a Bayesian odds $\ln\Odds_{C/R}\ge10$ to be the threshold for a confident association, although we emphasise that defining a threshold for confidence is subjective. A $\ln\Odds_{C/R} = 10$ implies that a common origin is $\approx 22,000$ times more likely than a random coincidence\footnote{We note the odds as calculated by \citet{ashton18} between GRB170817A and GW170817 is $\ln\Odds_{C/R}\gtrsim 12$, however this is not directly comparable to our results as they do not include the SNR dependence in the gravitational-wave trigger rate when calculating the prior odds.}. 
Our calculated Bayesian odds for a \grb-like neutron star binary in O3 sensitivity are summarised in Table.~\ref{tab:bfs}.

For the binary neutron star merger injected into the two (three) detector network at O3 sensitivity, the $\ln\Odds_{C/R}$ is $7.7^{+3.7}_{-3.9}$ ($8.3^{+3.7}_{-3.9}$), indicating that had two LIGO detectors been online at the time of GRB211211A, we would likely not have confidently associated the electromagnetic counterpart to the gravitational-wave signal (although this depends on the choice of threshold and prior odds). With O4 sensitivity, the binary neutron star $\ln\Odds_{C/R} = 29.1^{+3.7}_{-3.9}$, implying this would be a confident multi-messenger detection, despite the gravitational-wave signal being below the conventional threshold. Not only would such a detection have enabled the plethora of analyses similar to GW170817, but also questions unique to this event, perhaps unravelling the mystery of gamma-ray burst extended emission---we discuss this further in Section~\ref{sec:conclusion}. 

By contrast to the binary neutron star mergers, the odds for a neutron star-black hole merger with the two (three) detector network, is $\ln\Odds_{C/R}=33.3^{+3.7}_{-3.9}$ ($33.6^{+3.7}_{-3.9}$) at O3 sensitivity, implying we would confidently associate the electromagnetic counterpart to the gravitational-wave signal. We note that the neutron star-black hole gravitational-wave signal would also be below the conventional threshold for a network detection but would have a SNR$\gtrsim 8$ in the Livingston detector.  
\begin{table}
    \centering
    \begin{tabular}{|c|cc|}
        \hline
         & BNS & NSBH \\
         & $\ln{\mathcal{O}_{C/R}}$ & $\ln{\mathcal{O}_{C/R}}$\\
         \hline
    HL O3 & $7.7^{+3.7}_{-3.9}$ & $33.3^{+3.7}_{-3.9}$ \\[0.1cm]
    HLV O3 & $8.3^{+3.7}_{-3.9}$  & $33.6^{+3.7}_{-3.9}$ \\[0.1cm]
         \hline
    \end{tabular}
    \caption{Natural log of the common-to-random odds $\ln{\mathcal{O_{C/R}}}$ for a \grb{}-like binary neutron star merger (left column) and neutron star-black hole merger (right column) with a two and three gravitational-wave detector network at O3 sensitivity. The odds for both BNS and NSBH at O4 sensitivity are well above the threshold of detection.}
    \label{tab:bfs}
\end{table}

\section{Multi-messenger observation rates}
\grb{} may have been a missed opportunity---how do we derive optimal observing strategies to ensure we do not miss another multimessenger compact binary merger? To wit, we calculate the expected rate of coincident gravitational-wave and electromagnetic observations of mergers using the coincident detection criteria. 

We simulate binary neutron star and neutron star-black hole mergers isotropically distributed in the sky with inclination angles uniformly drawn from $\theta_{JN}\in\left[0^\circ,10^{\circ}\right]$, and intrinsic parameters $m_{\textrm{NS}}=1.4 M_{\odot}$, $m_{\textrm{BH}}=4.5 M_{\odot}$, $\chi_{\textrm{BH}} = 0.85$, $\chi_{\textrm{NS}}=0.01$, $R_{\textrm{NS}}=\unit[12]{km}$. Here, the subscripts NS and BH denote neutron star and black hole, respectively, and $M$, $R$, and $\chi$ are the objects' mass, radius, and in-plane dimensionless spin.
These parameters are specifically chosen as they are more likely to produce detectable prompt gamma-ray burst emission and kilonova than, say, mergers with higher inclination angle or more massive black holes.
In reality, only some fraction of neutron star mergers will be able to launch a jet or disrupt sufficient matter to produce a detectable kilonova~\citeg{sarin22_linking}. 
We expect this to have a small effect on the rates of detectable emission from binary neutron star mergers. However, the effect may be significantly larger for neutron star-black hole mergers, although the size of the effect is largely unknown given such system's unknown black hole mass and spin distribution.

We calculate the coincident odds for both binary neutron star and neutron star-black hole mergers with prompt gamma-ray and host galaxy observations as a function of distance for O4 and O5 (i.e., A+) projected sensitivities\footnote{Sensitivity curves are taken from \url{https://dcc.ligo.org/LIGO-T2000012/public}~\citep{abbott20_LRR}} with simulated Gaussian noise. 
To investigate optimal observing strategies, we inject our signals into two different detector networks: 1) a network of Livingston and Virgo, and 2) only Livingston. Note that these are proxy networks; our quantitative result would be largely unchanged if we, for example, used a single Virgo or KAGRA detector---the important aspect here is the sensitivity of the instruments, not their location. We note that at O5 sensitivity, Virgo is projected to be $\approx 60\%$ as sensitive as the Livingston detector. As in the previous section, we calculate the prior odds using Eq.~\ref{eq:priorodds}, with a gamma-ray trigger rate of $\unit[2]{day^{-1}}$ and a gravitational-wave trigger rate which is proportional to the signal-to-noise ratio.

\begin{figure*}
    \centering
    \includegraphics[width=1.95\columnwidth]{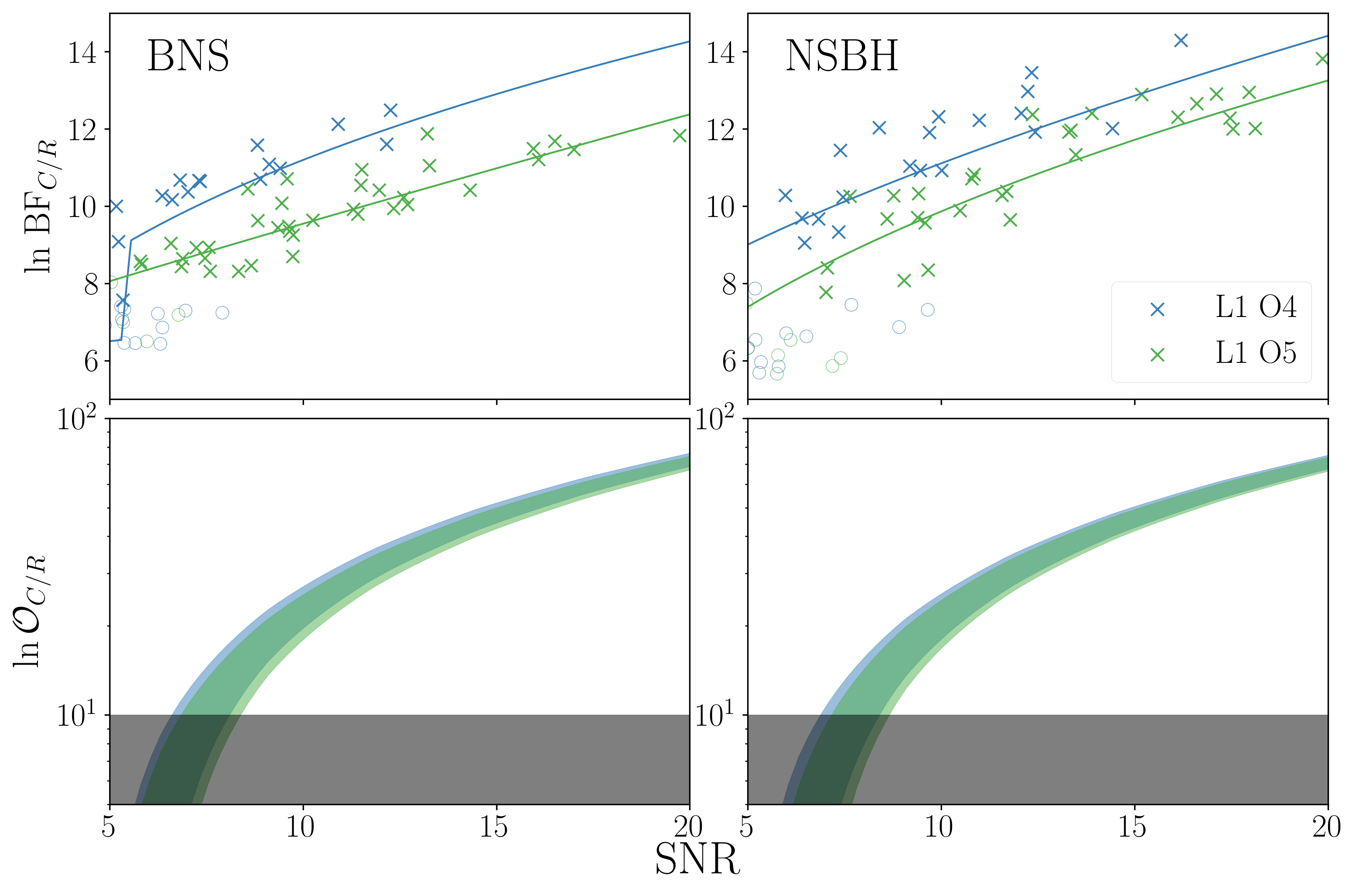}
    \caption{$\ln\mathrm{BF}_{C/R}$ ($\ln\Odds_{C/R}$) as a function of gravitational-wave SNR in the top (bottom) row for the binary neutron star (left panel) and neutron star-black hole (right panel) injections for a single Livingston detector operating at O4 (blue crosses) and O5 (green crosses) sensitivity. The black shaded region indicates $\ln\Odds_{C/R} <=10$ i.e., the region where one cannot confidently claim a multi-messenger detection. The faint circles illustrate injections where the $t_{c}$ posterior is not more informative than the prior, we consider these injections to be undetectable regardless of the $\ln\Odds_{C/R}$.}
    \label{fig:observingscenarios}
\end{figure*}

In Fig.~\ref{fig:observingscenarios}, we plot the $\ln\mathrm{BF}_{C/R}$ as a function of gravitational-wave SNR in the top row for the binary neutron star (left panel) and neutron star-black hole (right panel) injections for a single Livingston detector operating at O4 (blue crosses) and O5 (green crosses) sensitivity. The faint circles illustrate injections where the $t_{c}$ posterior $90\%$ credible interval is larger than $\unit[0.2]{s}$ and therefore the gravitational-wave analysis is not more informative than the prior. We do not consider these events coincident detections regardless of whether their $\ln\Odds_{C/R} \ge 10$. In the bottom row, we show the $\ln\Odds_{C/R}$ as a function of gravitational-wave SNR created by fitting a curve through the $\ln\mathrm{BF}_{C/R}$ shown in the top row and using Eq.~\ref{eq:priorodds} for the prior odds. We note that we fit a curve through $\ln\mathrm{BF}_{C/R}$ as creating a smooth $\ln\Odds_{C/R}$ vs. SNR curve would require many injections to account for the stochasticity of $\ln\mathrm{BF}_{C/R}$, which is computationally expensive and beyond the scope for this paper. In the bottom panel we also shade the region below $\ln\Odds_{C/R} = 10$, i.e., the region where we would not claim a confident multi-messenger detection. 

We find that across both networks, sensitivities and binary neutron star or neutron star-black hole mergers, the effective SNR where the gravitational-wave signal can be confidently associated to the gamma-ray counterpart is $\approx 7$ in a single gravitational-wave detector. This is lower than the conventional single detector threshold of $\rm{SNR} = 8$. That is, even though the gravitational-wave signal alone is not significant enough to claim a detection, the gamma-ray counterpart provides the necessary confidence for a multi-messenger detection. 


The above analysis assumes fixed masses, spins, and inclination angles. In reality, these properties dictate whether a neutron star binary can launch a jet that can produce gamma-ray emission and whether that jet is beamed at the observer. Refining this analysis, we take a realistic mass and spin distribution of neutron star binaries from an isolated binary evolution simulation (the so-called ``no pulsational pair instability supernova model'' in~\citet{Broekgaarden2021}) and calculate the rate of joint gravitational-wave signals and gamma-ray bursts out to a distance of $\unit[600]{Mpc}$ for binary neutron star mergers and $\unit[1000]{Mpc}$ for neutron star-black hole mergers for the same networks and sensitivities as above. We also account for whether a given binary can launch a jet and whether the gamma-ray burst is detectable using beaming and jet-launching fractions outlined in~\citet{sarin22_linking}.

\begin{table}
    \centering
    \begin{tabular}{|c|cccc|}
        \hline
         & BNS & BNS & NSBH & NSBH  \\
         & $\mathcal{F}$ & $\unit[\mathcal{R}]{[yr^{-1}]}$ & $\mathcal{F}$ & $\unit[\mathcal{R}]{[yr^{-1}]}$ \\
         \hline
         L O4 & 0.11 &  $\bnsrate$ & $\nsbhfrac$ &  $\nsbhrate$ \\[0.1cm]
         LV O4 & 0.10 & $0.21^{+7.8}_{-0.2}$ & 0.1 & $0.09^{+1.98}_{-0.09}$\\[0.1cm]
         L O5 & 0.34 &  $\bnsplusrate$ & $\nsbhplusfrac$ & $\nsbhplusrate$ \\[0.1cm]
         LV O5 & 0.37 &  $0.78^{+29.1}_{-0.76}$ & 0.34 &  $0.31^{+6.5}_{-0.31}$ \\[0.1cm]

         \hline
    \end{tabular}
    \caption{Predicted fraction $\mathcal{F}$ and rate $\mathcal{R}$ of binary neutron star (BNS) and neutron star-black hole (NSBH) binaries with confident coincident gravitational-wave and electromagnetic observations. We consider networks of two detectors consisting of LIGO-Livingston and Virgo (LV) and just Livingston (L) operating at O4 (i.e., design) and O5 (i.e., A+) sensitivities. The fraction of events is calculated assuming a luminosity distance $d_L\le\unit[600]{Mpc}$ for BNS and $d_L\le\unit[1000]{Mpc}$ for NSBH.}
    \label{tab:rates}
\end{table}

Our key results are summarised in Tab.~\ref{tab:rates}. The first column shows the fraction $\mathcal{F}$ of detected short gamma-ray bursts caused by binary neutron star mergers at luminosity distance $d_L\le\unit[600]{Mpc}$ that would be confident ($\ln\mathcal{O}_{C/R}\ge10$) multi-messenger detections for our different network configurations. 
For example, for a single LIGO-Livingston observatory operating at O4 sensitivity (L O4), approximately 11\% of short gamma-ray bursts from binary neutron stars within $\unit[600]{Mpc}$ would be confident multi-messenger detections. This number increases to 34\% using O5 (i.e., A+) sensitivity. Interestingly, this number does not change appreciably when adding another gravitational-wave detector, a direct consequence of the odds being predominantly informed by the measurement of the merger time, which is largely insensitive to the number of detectors in a network. 

This fraction translates into a rate of joint gravitational-wave and gamma-ray burst detections of $\unit[\bnsrate]{yr^{-1}}$ and $\unit[\bnsplusrate]{yr^{-1}}$ at O4 and O5 sensitivity, out to a horizon distance of $\unit[600]{Mpc}$ with the uncertainties coming from the uncertainty on the merger rate, beaming, and jet-launching fractions. 

We also calculate the fraction and rate for neutron star-black hole mergers of $\nsbhfrac$ ($\nsbhplusfrac$) and $\unit[\nsbhrate]{yr^{-1}}$ ($\unit[\nsbhplusrate]{yr^{-1}}$) at O4 (O5) sensitivity, respectively. These rates are much lower, a consequence of the smaller jet-launching fraction of neutron star-black hole mergers~\citeg{sarin22_linking}.
\section{Conclusion}\label{sec:conclusion}
\grb{} may have been caused by a binary neutron star or neutron star-black hole merger at $\approx\unit[350]{Mpc}$. If the latter, ground-based gravitational-wave observatories with their most recent (i.e., third-observing run; O3) sensitivity would have detected this multimessenger event if a single observatory were operating. This would have been a remarkable outcome: the gravitational-wave observations would have given us both the masses of the progenitors and the spin of the black hole. We note that, as with all measurements with gravitational waves, the chirp mass would have been best measured, and constraints on the component masses and mass ratio would have been weak.
This information, the kilonova observations, and numerical-relativity modelling of neutron star-black hole mergers, would have provided stringent constraints on the neutron star equation of state~\citeg{Coughlin2019}, and informed our understanding of r-process nucleosynthesis~\citeg{Metzger2019, Foucart2018}. 
Alternatively, if \grb{} was produced in the merger of a binary neutron star, a single gravitational-wave observatory operating at the proposed O4 sensitivity, will be able to make a confident multi-messenger detection. This is despite the gravitational-wave signal itself being below the conventional threshold for detection, with the gamma-ray burst providing the necessary confidence. Such a discovery would have shed light on the nature of the central engine in \grb{}.

With the imminent sensitivity improvement of gravitational-wave observatories, binary neutron star mergers with electromagnetic counterparts will be observable in gravitational waves at $\lesssim\unit[350]{Mpc}$, even with only a single observatory operating. We show this will be possible with the proposed O4 sensitivity, due to begin observing in early 2023 \citep{abbott20_LRR}. 
We find that the rate of joint gamma-ray and gravitational wave detections do not change significantly whether one gravitational-wave detector remains online or more. This is a direct consequence of the odds $\Odds_{C/R}$ being predominantly informed by the measurement of the time of coalescence. This motivates the question whether one single gravitational-wave detector should remain online at all times to not miss opportunities like \grb{}. 

We advocate that once gravitational-wave detectors reach O5 sensitivity, at least one detector remain online at all times. This will ensure that at least $\bnsplusrate$ ($\nsbhplusrate$) per year of gamma-ray bursts produced by binary neutron star (neutron star-black hole mergers) out to a distance of $\unit[600]{Mpc}$ ($\unit[1000]{Mpc}$) can be associated to their often sub-threshold gravitational-wave signal enabling the rich promise of multi-messenger astronomy, while other detectors in the network perform maintenance or upgrades. Ignoring uncertainties from the beaming and jet-launching fraction, this would mean that $\approx 34\%$ ($35\%$) of short gamma-ray bursts within $\unit[600]{Mpc}$ ($\unit[1000]{Mpc}$) produced by a binary 
neutron star (neutron star-black hole) would yield a confident gravitational-wave detection with a single gravitational-wave detector. We emphasize that this calculation assumes that the gamma-ray burst redshift is measured, which therefore requires gamma-ray burst triggers continue to be followed up independently and produce a successful identification of the host galaxy and or optical counterpart.

Even though the gravitational-wave signals may be sub-threshold, the potential science output of multimessenger observations as described here is immense. For example:
\begin{itemize}
    \item \textit{Hubble constant measurements.} The 90\% credible-interval posterior width on the luminosity distance measured with gravitational waves from a single-detector binary neutron star merger is $\unit[200]{Mpc}$ (assuming conservatively, that the observation of a gamma-ray burst does not provide any measurement of the inclination angle). This implies that the sub-threshold gravitational-wave signals with host-galaxy redshifts can be used for accurate, distance-ladder-free inferences of the Hubble constant~\citep{schutz86,abbott17_GW170817_hubble}. For example, assuming \grb{} was a binary neutron star merger, a single gravitational-wave detector operating at O4 sensitivity would have produced a Hubble constant measurement of $H_{0} = \unit[{74.72}_{-11.64}^{+21.20}]{km\,s^{-1}\,Mpc^{-1}}$ without including any constraint on the viewing angle from the gamma-ray burst. If the gamma-ray burst measurement of the viewing angle was included, the Hubble constant measurement would have been $H_{0} = \unit[{62.12}_{-3.96}^{+5.09}]{{km\,s^{-1}\,Mpc^{-1}}}$.
    \item \textit{Tests of gravity.} The gravitational-wave uncertainty on the time of coalescence, even in the case of a single-detector observation with matched filter signal-to-noise ratio $\approx 7$, is $\lesssim\unit[0.1]{s}$. Comparing this time with the prompt emission will therefore allow constraints on the speed of gravity, akin to what was achieved with multimessenger observations of GW170817~\citep{abbott17_GW170817_mma}
    \item \textit{Equation of state.} While the tidal deformability of neutron star mergers cannot be accurately measured from such sub-threshold events (the posterior distribution is similar to the prior in our simulations), knowledge of the progenitor masses does allow for modelled calculations of the ejecta mass that can be compared to e.g., the kilonova observations to provide constraints on the equation of state~\citeg{Coughlin2019, Nicholl2021}. Electromagnetic observations also provide constraints on the fate of binary neutron star mergers~\citeg{sarin21_review}, which can be combined with measurements of the progenitor masses to constrain the maximum neutron star mass~\citeg{Margalit2017}. The $90\%$ credible-interval posterior width of the progenitor chirp mass and mass ratio in our single-detector O4 Livingston observation for a \grb-like binary neutron star is  $\lesssim\unit[0.0023]{M_\odot}$ and $\lesssim 0.47$, respectively. 
    \item \textit{Gamma-ray burst physics.} Comparing the time of coalescence with the prompt gamma-ray emission also allows for studies into the physics of the the delay time between the merger, and the launch and propagation of the jet. Furthermore, given an approximate distance to the source is known from the host galaxy, an approximate viewing angle can be determined from the gravitational-wave posterior distribution of the inclination angle. In the case of a single-detector Livingston observation in O4, this inclination angle can be determined with an uncertainty of $20^\circ$ from the gravitational-wave data alone, implying the physics of jet opening angles and angular energy distributions can be studied~\citep[e.g.,][]{mandel18, Lamb2022, Nativi2022}.
\end{itemize}

The above laundry list provides a somewhat narrow view of the potential science for gravitational-wave multimessenger astronomy by only focusing on neutron star mergers. Coincident gravitational-wave and electromagnetic observations from nearby (i.e., galactic) supernovae would provide a rich cornucopia of science, however the event rate of such events is currently believed to be $\lesssim\unit[1]{/(50\,yr)}$~\citep{Tammann1994}, implying one should minimise the downtime of gravitational-wave observatories and continue to improve the sensitivity of the instruments. Of course, there are other potentially rare sources of gravitational waves that could be accompanied by electromagnetic observations such as giant magnetar flares, pulsar glitches, and fast-radio bursts~\citep{Macquet2021, lvk_frb2022, yang22}. 

While for the above reasons it is worth considering whether one detector with sufficient sensitivity should always remain in sky-watch mode, such a proposal needs to be tempered by commissioning considerations. Furthermore, this ignores the consideration for following up gravitational-wave observations to find kilonovae, which rely on multiple detectors for a small sky map. Therefore, any consideration also needs to account for the difficulty of finding kilonovae in follow-up to a short gamma-ray burst trigger as opposed to a gravitational-wave trigger. 
This is perhaps simple to answer currently as, due to the size of the visible volume in gravitational waves and beaming effects, the detectable kilonova rate will be larger than the detectable gamma-ray burst rate~\citeg{Metzger2012, kelley13}. 
However, as gravitational-wave interferometers near third-generation sensitivities, this will quickly change, at which point it will also become important to consider the sensitivity of gamma-ray observatories, which will become the limiting factor for joint detections. 
Such a proposal must also weigh the affect of this sky-watch mode on science from binary black hole mergers. Although several studies have looked at the aspect of low-latency detection for binary black holes with a single detector~\citeg{callister17, davies22}, parameter estimation and astrophysical inference (e.g., how measurements of spin precession, eccentricity, higher order modes are affected by only having a single detector, or the implications for population-level analyses) have not been studied in detail.

\section*{Acknowledgements}
We are grateful to Gareth Cabourn Davies, Steve Fairhurst, Ilya Mandel, David Ottaway, Fiona Panther, Simon Stevenson, Eric Thrane, Ariel Goobar, and Michael Coughlin for valuable discussions.
This work was supported through Australian Research Council (ARC) Centre of Excellence CE170100004, and ARC Discovery Project DP220101610. N.S is supported by a Nordita Fellowship, Nordita is supported in part by NordForsk. This work was performed in part at Aspen Center for Physics, which is supported by National Science Foundation grant PHY-1607611. This work was also performed on the OzSTAR national facility at Swinburne University of Technology. The OzSTAR program receives funding in part from the Astronomy National Collaborative Research Infrastructure Strategy (NCRIS) allocation provided by the Australian Government. This material is based upon work supported by NSF's LIGO Laboratory which is a major facility fully funded by the National Science Foundation.
\section*{Data Availability}
No data was used in this manuscript.

\bibliographystyle{mnras}
\bibliography{bib}

\begin{thebibliography}{}
\makeatletter
\relax
\def\mn@urlcharsother{\let\do\@makeother \do\$\do\&\do\#\do\^\do\_\do\%\do\~}
\def\mn@doi{\begingroup\mn@urlcharsother \@ifnextchar [ {\mn@doi@}
  {\mn@doi@[]}}
\def\mn@doi@[#1]#2{\def\@tempa{#1}\ifx\@tempa\@empty \href
  {http://dx.doi.org/#2} {doi:#2}\else \href {http://dx.doi.org/#2} {#1}\fi
  \endgroup}
\def\mn@eprint#1#2{\mn@eprint@#1:#2::\@nil}
\def\mn@eprint@arXiv#1{\href {http://arxiv.org/abs/#1} {{\tt arXiv:#1}}}
\def\mn@eprint@dblp#1{\href {http://dblp.uni-trier.de/rec/bibtex/#1.xml}
  {dblp:#1}}
\def\mn@eprint@#1:#2:#3:#4\@nil{\def\@tempa {#1}\def\@tempb {#2}\def\@tempc
  {#3}\ifx \@tempc \@empty \let \@tempc \@tempb \let \@tempb \@tempa \fi \ifx
  \@tempb \@empty \def\@tempb {arXiv}\fi \@ifundefined
  {mn@eprint@\@tempb}{\@tempb:\@tempc}{\expandafter \expandafter \csname
  mn@eprint@\@tempb\endcsname \expandafter{\@tempc}}}

\bibitem[\protect\citeauthoryear{{Aasi} et~al.}{{Aasi} et~al.}{2015}]{LIGO}
{Aasi} J.,  et~al., 2015, \mn@doi [CQG] {10.1088/0264-9381/32/7/074001}, 32,
  074001

\bibitem[\protect\citeauthoryear{{Abbott} et~al.,}{{Abbott}
  et~al.}{2017a}]{abbott17_GW170817_hubble}
{Abbott} B.~P.,  et~al., 2017a, \mn@doi [\nat] {10.1038/nature24471}, \href
  {https://ui.adsabs.harvard.edu/abs/2017Natur.551...85A} {551, 85}

\bibitem[\protect\citeauthoryear{{Abbott} et~al.,}{{Abbott}
  et~al.}{2017b}]{abbott17_GW170817_mma}
{Abbott} B.~P.,  et~al., 2017b, \mn@doi [\apjl] {10.3847/2041-8213/aa91c9},
  \href {https://ui.adsabs.harvard.edu/abs/2017ApJ...848L..12A} {848, L12}

\bibitem[\protect\citeauthoryear{{Abbott} et~al.,}{{Abbott}
  et~al.}{2020}]{abbott20_LRR}
{Abbott} B.~P.,  et~al., 2020, \mn@doi [Living Reviews in Relativity]
  {10.1007/s41114-020-00026-9}, \href
  {https://ui.adsabs.harvard.edu/abs/2020LRR....23....3A} {23, 3}

\bibitem[\protect\citeauthoryear{{Abbott}, {Abbott}, {Acernese}, {Ackley}  \&
  et al.}{{Abbott} et~al.}{2021a}]{gwtc3}
{Abbott} R.,  {Abbott} T.~D.,  {Acernese} F.,  {Ackley} K.,   et al. 2021a,
  arXiv e-prints, \href {https://ui.adsabs.harvard.edu/abs/2021arXiv211103606T}
  {p. arXiv:2111.03606}

\bibitem[\protect\citeauthoryear{{Abbott} et~al.,}{{Abbott}
  et~al.}{2021b}]{abbott22_gwtc3pop}
{Abbott} R.,  et~al., 2021b, arXiv e-prints, \href
  {https://ui.adsabs.harvard.edu/abs/2021arXiv211103634T} {p. arXiv:2111.03634}

\bibitem[\protect\citeauthoryear{{Abbott} et~al.,}{{Abbott}
  et~al.}{2022a}]{lvk_frb2022}
{Abbott} B.~P.,  et~al., 2022a, arXiv e-prints, \href
  {https://ui.adsabs.harvard.edu/abs/2022arXiv220312038T} {p. arXiv:2203.12038}

\bibitem[\protect\citeauthoryear{{Abbott}, {Abbott}, {Acernese}, {Ackley}  \&
  et al.}{{Abbott} et~al.}{2022b}]{Abbott2022grbb}
{Abbott} R.,  {Abbott} T.~D.,  {Acernese} F.,  {Ackley} K.,   et al. 2022b,
  \mn@doi [\apj] {10.3847/1538-4357/ac532b}, \href
  {https://ui.adsabs.harvard.edu/abs/2022ApJ...928..186A} {928, 186}

\bibitem[\protect\citeauthoryear{{Acernese} et~al.}{{Acernese}
  et~al.}{2015}]{Virgo}
{Acernese} F.,  et~al., 2015, \mn@doi [] {10.1088/0264-9381/32/2/024001}, 32,
  024001

\bibitem[\protect\citeauthoryear{{Aghanim} et~al.,}{{Aghanim}
  et~al.}{2020}]{aghanim20}
{Aghanim} N.,  et~al., 2020, \mn@doi [\aap] {10.1051/0004-6361/201833910},
  \href {https://ui.adsabs.harvard.edu/abs/2020A&A...641A...6P} {641, A6}

\bibitem[\protect\citeauthoryear{{Akutsu}, {Ando}, {Arai}  \& et al.}{{Akutsu}
  et~al.}{2019}]{KAGRA}
{Akutsu} T.,  {Ando} M.,  {Arai} K.,   et al. 2019, \mn@doi [Nature Astronomy]
  {10.1038/s41550-018-0658-y}, \href
  {https://ui.adsabs.harvard.edu/abs/2019NatAs...3...35K} {3, 35}

\bibitem[\protect\citeauthoryear{{Ashton} et~al.,}{{Ashton}
  et~al.}{2018}]{ashton18}
{Ashton} G.,  et~al., 2018, \mn@doi [\apj] {10.3847/1538-4357/aabfd2}, \href
  {https://ui.adsabs.harvard.edu/abs/2018ApJ...860....6A} {860, 6}

\bibitem[\protect\citeauthoryear{{Ashton} et~al.,}{{Ashton}
  et~al.}{2019}]{ashton19}
{Ashton} G.,  et~al., 2019, \mn@doi [\apjs] {10.3847/1538-4365/ab06fc}, \href
  {https://ui.adsabs.harvard.edu/abs/2019ApJS..241...27A} {241, 27}

\bibitem[\protect\citeauthoryear{{Bartos}, {Veske}, {Keivani}, {M{\'a}rka},
  {Countryman}, {Blaufuss}, {Finley}  \& {M{\'a}rka}}{{Bartos}
  et~al.}{2019}]{bartos19}
{Bartos} I.,  {Veske} D.,  {Keivani} A.,  {M{\'a}rka} Z.,  {Countryman} S.,
  {Blaufuss} E.,  {Finley} C.,   {M{\'a}rka} S.,  2019, \mn@doi [\prd]
  {10.1103/PhysRevD.100.083017}, \href
  {https://ui.adsabs.harvard.edu/abs/2019PhRvD.100h3017B} {100, 083017}

\bibitem[\protect\citeauthoryear{{Broekgaarden}, {Berger}, {Neijssel},
  {Vigna-G{\'o}mez}  \& et al.}{{Broekgaarden} et~al.}{2021}]{Broekgaarden2021}
{Broekgaarden} F.~S.,  {Berger} E.,  {Neijssel} C.~J.,  {Vigna-G{\'o}mez} A.,
  et al. 2021, \mn@doi [\mnras] {10.1093/mnras/stab2716}, \href
  {https://ui.adsabs.harvard.edu/abs/2021MNRAS.508.5028B} {508, 5028}

\bibitem[\protect\citeauthoryear{{Cabourn Davies} \& Harry}{{Cabourn Davies} \&
  Harry}{2022}]{davies22}
{Cabourn Davies} G.~S.,  Harry I.~W.,  2022, \mn@doi [Class. Quant. Grav.]
  {10.1088/1361-6382/ac8862}, 39, 215012

\bibitem[\protect\citeauthoryear{{Callister}, {Kanner}, {Massinger},
  {Dhurandhar}  \& {Weinstein}}{{Callister} et~al.}{2017}]{callister17}
{Callister} T.~A.,  {Kanner} J.~B.,  {Massinger} T.~J.,  {Dhurandhar} S.,
  {Weinstein} A.~J.,  2017, \mn@doi [Classical and Quantum Gravity]
  {10.1088/1361-6382/aa7a76}, \href
  {https://ui.adsabs.harvard.edu/abs/2017CQGra..34o5007C} {34, 155007}

\bibitem[\protect\citeauthoryear{{Cannon}, {Caudill}, {Chan}, {Cousins}  \& et
  al.}{{Cannon} et~al.}{2021}]{Cannon2021}
{Cannon} K.,  {Caudill} S.,  {Chan} C.,  {Cousins} B.,   et al. 2021, \mn@doi
  [SoftwareX] {10.1016/j.softx.2021.100680}, \href
  {https://ui.adsabs.harvard.edu/abs/2021SoftX..1400680C} {14, 100680}

\bibitem[\protect\citeauthoryear{{Clark}, {Evans}, {Fairhurst}, {Harry}  \& et
  al.}{{Clark} et~al.}{2015}]{Clark2015}
{Clark} J.,  {Evans} H.,  {Fairhurst} S.,  {Harry} I.~W.,   et al. 2015,
  \mn@doi [\apj] {10.1088/0004-637X/809/1/53}, \href
  {https://ui.adsabs.harvard.edu/abs/2015ApJ...809...53C} {809, 53}

\bibitem[\protect\citeauthoryear{{Colombo}, {Salafia}, {Gabrielli}, {Ghirlanda}
   \& et al.}{{Colombo} et~al.}{2022}]{Colombo2022}
{Colombo} A.,  {Salafia} O.~S.,  {Gabrielli} F.,  {Ghirlanda} G.,   et al.
  2022, \mn@doi [\apj] {10.3847/1538-4357/ac8d00}, \href
  {https://ui.adsabs.harvard.edu/abs/2022ApJ...937...79C} {937, 79}

\bibitem[\protect\citeauthoryear{{Coughlin}, {Dietrich}, {Margalit}  \&
  {Metzger}}{{Coughlin} et~al.}{2019}]{Coughlin2019}
{Coughlin} M.~W.,  {Dietrich} T.,  {Margalit} B.,   {Metzger} B.~D.,  2019,
  \mn@doi [\mnras] {10.1093/mnrasl/slz133}, \href
  {https://ui.adsabs.harvard.edu/abs/2019MNRAS.489L..91C} {489, L91}

\bibitem[\protect\citeauthoryear{{D'Ai}, {Ambrosi}, {D'Elia}, {Gropp}  \& et
  al.}{{D'Ai} et~al.}{2021}]{D'Ai2021}
{D'Ai} A.,  {Ambrosi} E.,  {D'Elia} V.,  {Gropp} J.~D.,   et al. 2021, GRB
  Coordinates Network, \href
  {https://ui.adsabs.harvard.edu/abs/2021GCN.31202....1D} {31202, 1}

\bibitem[\protect\citeauthoryear{{Fermi GBM Team}}{{Fermi GBM
  Team}}{2021}]{FermiGBMTeam2021}
{Fermi GBM Team} 2021, GRB Coordinates Network, \href
  {https://ui.adsabs.harvard.edu/abs/2021GCN.31201....1F} {31201, 1}

\bibitem[\protect\citeauthoryear{{Fong}, {Berger}, {Margutti}  \&
  {Zauderer}}{{Fong} et~al.}{2015}]{Fong2015}
{Fong} W.,  {Berger} E.,  {Margutti} R.,   {Zauderer} B.~A.,  2015, \mn@doi
  [\apj] {10.1088/0004-637X/815/2/102}, \href
  {https://ui.adsabs.harvard.edu/abs/2015ApJ...815..102F} {815, 102}

\bibitem[\protect\citeauthoryear{Foucart, Hinderer  \& Nissanke}{Foucart
  et~al.}{2018}]{Foucart2018}
Foucart F.,  Hinderer T.,   Nissanke S.,  2018, \mn@doi [Phys. Rev. D]
  {10.1103/PhysRevD.98.081501}, 98, 081501

\bibitem[\protect\citeauthoryear{{Gao}, {Lei}  \& {Zhu}}{{Gao}
  et~al.}{2022}]{Gao2022}
{Gao} H.,  {Lei} W.-H.,   {Zhu} Z.-P.,  2022, arXiv e-prints, \href
  {https://ui.adsabs.harvard.edu/abs/2022arXiv220505031G} {p. arXiv:2205.05031}

\bibitem[\protect\citeauthoryear{{Gompertz}, {Ravasio}, {Nicholl}, {Levan}  \&
  et al.}{{Gompertz} et~al.}{2022}]{Gompertz2022}
{Gompertz} B.~P.,  {Ravasio} M.~E.,  {Nicholl} M.,  {Levan} A.~J.,   et al.
  2022, arXiv e-prints, \href
  {https://ui.adsabs.harvard.edu/abs/2022arXiv220505008G} {p. arXiv:2205.05008}

\bibitem[\protect\citeauthoryear{{Kelley}, {Mandel}  \&
  {Ramirez-Ruiz}}{{Kelley} et~al.}{2013}]{kelley13}
{Kelley} L.~Z.,  {Mandel} I.,   {Ramirez-Ruiz} E.,  2013, \mn@doi [\prd]
  {10.1103/PhysRevD.87.123004}, \href
  {https://ui.adsabs.harvard.edu/abs/2013PhRvD..87l3004K} {87, 123004}

\bibitem[\protect\citeauthoryear{{Lamb}, {Nativi}, {Rosswog}, {Kann}  \& et
  al.}{{Lamb} et~al.}{2022}]{Lamb2022}
{Lamb} G.~P.,  {Nativi} L.,  {Rosswog} S.,  {Kann} D.~A.,   et al. 2022, arXiv
  e-prints, \href {https://ui.adsabs.harvard.edu/abs/2022arXiv220109796L} {p.
  arXiv:2201.09796}

\bibitem[\protect\citeauthoryear{{Macquet}, {Bizouard}, {Burns}, {Christensen}
  \& et al.}{{Macquet} et~al.}{2021}]{Macquet2021}
{Macquet} A.,  {Bizouard} M.~A.,  {Burns} E.,  {Christensen} N.,   et al. 2021,
  \mn@doi [\apj] {10.3847/1538-4357/ac0efd}, \href
  {https://ui.adsabs.harvard.edu/abs/2021ApJ...918...80M} {918, 80}

\bibitem[\protect\citeauthoryear{{Magee} et~al.,}{{Magee}
  et~al.}{2019}]{magee19}
{Magee} R.,  et~al., 2019, \mn@doi [\apjl] {10.3847/2041-8213/ab20cf}, \href
  {https://ui.adsabs.harvard.edu/abs/2019ApJ...878L..17M} {878, L17}

\bibitem[\protect\citeauthoryear{{Mandel}}{{Mandel}}{2018}]{mandel18}
{Mandel} I.,  2018, \mn@doi [\apjl] {10.3847/2041-8213/aaa6c1}, \href
  {https://ui.adsabs.harvard.edu/abs/2018ApJ...853L..12M} {853, L12}

\bibitem[\protect\citeauthoryear{{Margalit} \& {Metzger}}{{Margalit} \&
  {Metzger}}{2017}]{Margalit2017}
{Margalit} B.,  {Metzger} B.~D.,  2017, \mn@doi [\apjl]
  {10.3847/2041-8213/aa991c}, \href
  {https://ui.adsabs.harvard.edu/abs/2017ApJ...850L..19M} {850, L19}

\bibitem[\protect\citeauthoryear{{Metzger}}{{Metzger}}{2019}]{Metzger2019}
{Metzger} B.~D.,  2019, \mn@doi [Living Reviews in Relativity]
  {10.1007/s41114-019-0024-0}, \href
  {https://ui.adsabs.harvard.edu/abs/2019LRR....23....1M} {23, 1}

\bibitem[\protect\citeauthoryear{{Metzger} \& {Berger}}{{Metzger} \&
  {Berger}}{2012}]{Metzger2012}
{Metzger} B.~D.,  {Berger} E.,  2012, \mn@doi [\apj]
  {10.1088/0004-637X/746/1/48}, \href
  {https://ui.adsabs.harvard.edu/abs/2012ApJ...746...48M} {746, 48}

\bibitem[\protect\citeauthoryear{{Nativi}, {Lamb}, {Rosswog}, {Lundman}  \& et
  al.}{{Nativi} et~al.}{2022}]{Nativi2022}
{Nativi} L.,  {Lamb} G.~P.,  {Rosswog} S.,  {Lundman} C.,   et al. 2022,
  \mn@doi [\mnras] {10.1093/mnras/stab2982}, \href
  {https://ui.adsabs.harvard.edu/abs/2022MNRAS.509..903N} {509, 903}

\bibitem[\protect\citeauthoryear{{Nicholl}, {Margalit}, {Schmidt}, {Smith},
  {Ridley}  \& {Nuttall}}{{Nicholl} et~al.}{2021}]{Nicholl2021}
{Nicholl} M.,  {Margalit} B.,  {Schmidt} P.,  {Smith} G.~P.,  {Ridley} E.~J.,
  {Nuttall} J.,  2021, \mn@doi [\mnras] {10.1093/mnras/stab1523}, \href
  {https://ui.adsabs.harvard.edu/abs/2021MNRAS.505.3016N} {505, 3016}

\bibitem[\protect\citeauthoryear{{Nitz}, {Nielsen}  \& {Capano}}{{Nitz}
  et~al.}{2019}]{nitz19}
{Nitz} A.~H.,  {Nielsen} A.~B.,   {Capano} C.~D.,  2019, \mn@doi [\apjl]
  {10.3847/2041-8213/ab18a1}, \href
  {https://ui.adsabs.harvard.edu/abs/2019ApJ...876L...4N} {876, L4}

\bibitem[\protect\citeauthoryear{{Nitz}, {Dent}, {Davies}  \& {Harry}}{{Nitz}
  et~al.}{2020}]{nitz20}
{Nitz} A.~H.,  {Dent} T.,  {Davies} G.~S.,   {Harry} I.,  2020, \mn@doi [\apj]
  {10.3847/1538-4357/ab96c7}, \href
  {https://ui.adsabs.harvard.edu/abs/2020ApJ...897..169N} {897, 169}

\bibitem[\protect\citeauthoryear{{Patricelli}, {Bernardini}, {Mapelli},
  {D'Avanzo}  \& et al.}{{Patricelli} et~al.}{2022}]{Patricelli2022}
{Patricelli} B.,  {Bernardini} M.~G.,  {Mapelli} M.,  {D'Avanzo} P.,   et al.
  2022, \mn@doi [\mnras] {10.1093/mnras/stac1167}, \href
  {https://ui.adsabs.harvard.edu/abs/2022MNRAS.513.4159P} {513, 4159}

\bibitem[\protect\citeauthoryear{{Petrov}, {Singer}, {Coughlin}, {Kumar}  \& et
  al.}{{Petrov} et~al.}{2022}]{Petrov2022}
{Petrov} P.,  {Singer} L.~P.,  {Coughlin} M.~W.,  {Kumar} V.,   et al. 2022,
  \mn@doi [\apj] {10.3847/1538-4357/ac366d}, \href
  {https://ui.adsabs.harvard.edu/abs/2022ApJ...924...54P} {924, 54}

\bibitem[\protect\citeauthoryear{{Rastinejad} et~al.,}{{Rastinejad}
  et~al.}{2022}]{rastinejad22}
{Rastinejad} J.~C.,  et~al., 2022, arXiv e-prints, \href
  {https://ui.adsabs.harvard.edu/abs/2022arXiv220410864R} {p. arXiv:2204.10864}

\bibitem[\protect\citeauthoryear{{Romero-Shaw} et~al.,}{{Romero-Shaw}
  et~al.}{2020}]{romeroshaw20}
{Romero-Shaw} I.~M.,  et~al., 2020, \mn@doi [\mnras] {10.1093/mnras/staa2850},
  \href {https://ui.adsabs.harvard.edu/abs/2020MNRAS.499.3295R} {499, 3295}

\bibitem[\protect\citeauthoryear{{Sarin} \& {Lasky}}{{Sarin} \&
  {Lasky}}{2021}]{sarin21_review}
{Sarin} N.,  {Lasky} P.~D.,  2021, \mn@doi [General Relativity and Gravitation]
  {10.1007/s10714-021-02831-1}, \href
  {https://ui.adsabs.harvard.edu/abs/2021GReGr..53...59S} {53, 59}

\bibitem[\protect\citeauthoryear{{Sarin} \& {Lasky}}{{Sarin} \&
  {Lasky}}{2022}]{sarin22}
{Sarin} N.,  {Lasky} P.~D.,  2022, \mn@doi [\pasa] {10.1017/pasa.2022.1}, \href
  {https://ui.adsabs.harvard.edu/abs/2022PASA...39....7S} {39, e007}

\bibitem[\protect\citeauthoryear{{Sarin}, {Lasky}, {Vivanco}, {Stevenson},
  {Chattopadhyay}, {Smith}  \& {Thrane}}{{Sarin}
  et~al.}{2022}]{sarin22_linking}
{Sarin} N.,  {Lasky} P.~D.,  {Vivanco} F.~H.,  {Stevenson} S.~P.,
  {Chattopadhyay} D.,  {Smith} R.,   {Thrane} E.,  2022, \mn@doi [\prd]
  {10.1103/PhysRevD.105.083004}, \href
  {https://ui.adsabs.harvard.edu/abs/2022PhRvD.105h3004S} {105, 083004}

\bibitem[\protect\citeauthoryear{{Schutz}}{{Schutz}}{1986}]{schutz86}
{Schutz} B.~F.,  1986, \mn@doi [\nat] {10.1038/323310a0}, \href
  {https://ui.adsabs.harvard.edu/abs/1986Natur.323..310S} {323, 310}

\bibitem[\protect\citeauthoryear{{Speagle}}{{Speagle}}{2020}]{speagle20}
{Speagle} J.~S.,  2020, \mn@doi [\mnras] {10.1093/mnras/staa278}, \href
  {https://ui.adsabs.harvard.edu/abs/2020MNRAS.493.3132S} {493, 3132}

\bibitem[\protect\citeauthoryear{{Stachie} et~al.,}{{Stachie}
  et~al.}{2020}]{stachie20}
{Stachie} C.,  et~al., 2020, \mn@doi [Classical and Quantum Gravity]
  {10.1088/1361-6382/aba28a}, \href
  {https://ui.adsabs.harvard.edu/abs/2020CQGra..37q5001S} {37, 175001}

\bibitem[\protect\citeauthoryear{{Suvorov}, {Kuan}  \& {Kokkotas}}{{Suvorov}
  et~al.}{2022}]{Suvorov2022}
{Suvorov} A.~G.,  {Kuan} H.-J.,   {Kokkotas} K.~D.,  2022, arXiv e-prints,
  \href {https://ui.adsabs.harvard.edu/abs/2022arXiv220511112S} {p.
  arXiv:2205.11112}

\bibitem[\protect\citeauthoryear{{Tammann}, {Loeffler}  \&
  {Schroeder}}{{Tammann} et~al.}{1994}]{Tammann1994}
{Tammann} G.~A.,  {Loeffler} W.,   {Schroeder} A.,  1994, \mn@doi [\apjs]
  {10.1086/192002}, \href
  {https://ui.adsabs.harvard.edu/abs/1994ApJS...92..487T} {92, 487}

\bibitem[\protect\citeauthoryear{{Troja}, {Fryer}, {O'Connor}, {Ryan}  \& et
  al.}{{Troja} et~al.}{2022}]{Troja2022}
{Troja} E.,  {Fryer} C.~L.,  {O'Connor} B.,  {Ryan} G.,   et al. 2022, arXiv
  e-prints, \href {https://ui.adsabs.harvard.edu/abs/2022arXiv220903363T} {p.
  arXiv:2209.03363}

\bibitem[\protect\citeauthoryear{{Waxman}, {Ofek}  \& {Kushnir}}{{Waxman}
  et~al.}{2022}]{Waxman2022}
{Waxman} E.,  {Ofek} E.~O.,   {Kushnir} D.,  2022, arXiv e-prints, \href
  {https://ui.adsabs.harvard.edu/abs/2022arXiv220610710W} {p. arXiv:2206.10710}

\bibitem[\protect\citeauthoryear{{Willke}, {Aufmuth}, {Aulbert}, {Babak}  \& et
  al.}{{Willke} et~al.}{2002}]{Willke2002}
{Willke} B.,  {Aufmuth} P.,  {Aulbert} C.,  {Babak} S.,   et al. 2002, \mn@doi
  [Classical and Quantum Gravity] {10.1088/0264-9381/19/7/321}, \href
  {https://ui.adsabs.harvard.edu/abs/2002CQGra..19.1377W} {19, 1377}

\bibitem[\protect\citeauthoryear{{Xiao} et~al.,}{{Xiao} et~al.}{2022}]{xiao22}
{Xiao} S.,  et~al., 2022, arXiv e-prints, \href
  {https://ui.adsabs.harvard.edu/abs/2022arXiv220502186X} {p. arXiv:2205.02186}

\bibitem[\protect\citeauthoryear{Yang, Zhang, Ai, Liu, Wang, L\"u  \&
  Zhang}{Yang et~al.}{2022}]{yang22}
Yang J.,  Zhang B.-B.,  Ai S.,  Liu Z.-K.,  Wang X.,  L\"u H.-J.,   Zhang B.,
  2022, arXiv e-prints, \href
  {https://ui.adsabs.harvard.edu/abs/2022arXiv220412771Y} {p. arXiv:2204.12771}

\bibitem[\protect\citeauthoryear{{von Kienlin}, {Meegan}, {Paciesas}, {Bhat}
  \& et al.}{{von Kienlin} et~al.}{2020}]{vonKienlin2020}
{von Kienlin} A.,  {Meegan} C.~A.,  {Paciesas} W.~S.,  {Bhat} P.~N.,   et al.
  2020, \mn@doi [\apj] {10.3847/1538-4357/ab7a18}, \href
  {https://ui.adsabs.harvard.edu/abs/2020ApJ...893...46V} {893, 46}

\makeatother
\end{thebibliography}
\end{document}